\documentclass{article}
\newcommand\independent{\protect\mathpalette{\protect\independenT}{\perp}}
\AtBeginDocument{%
  \providecommand\BibTeX{{%
    \normalfont B\kern-0.5em{\scshape i\kern-0.25em b}\kern-0.8em\TeX}}}
\usepackage[accepted]{icml2024}
\usepackage{microtype}
\usepackage{tikz}
\usepackage{amsthm}
\usepackage{amsmath}
\usepackage{amsthm}
\usepackage{dsfont}
\usepackage{booktabs}
\usepackage{amsfonts}
\usepackage{mathrsfs}
\usepackage{graphicx}
\usepackage{subfigure}
\usepackage{color}
\usepackage{float}
\usepackage{algorithm,algorithmic}
\usepackage{multirow}
\usepackage{hyperref}
\usepackage{xcolor}
\usepackage[capitalize,noabbrev]{cleveref}
\newtheorem{assumption}{Assumption}

\newtheorem{theorem}{Theorem}

\newtheorem{lemma}[theorem]{Lemma}
\newtheorem{corollary}[theorem]{Corollary}

\allowdisplaybreaks
\DeclareMathOperator*{\argmin}{arg\,min}

\newcommand{\indep}{\perp \!\!\! \perp}

\icmltitlerunning{Long-Term Causal Effects of Long-Term Treatments}

\begin{document}
\twocolumn[
\icmltitle{Inferring the Long-Term Causal Effects of Long-Term Treatments
\\from Short-Term Experiments}



\icmlsetsymbol{equal}{*}

\begin{icmlauthorlist}
\icmlauthor{Allen Tran}{xxx}
\icmlauthor{Aurélien Bibaut}{yyy}
\icmlauthor{Nathan Kallus}{yyy,comp}
\end{icmlauthorlist}

\icmlaffiliation{xxx}{Netflix Inc., Los Angeles, USA}
\icmlaffiliation{yyy}{Netflix Inc., Los Gatos, USA}
\icmlaffiliation{comp}{Cornell University, New York, USA}

\icmlcorrespondingauthor{Allen Tran}{allent@netflix.com}

\icmlkeywords{Machine Learning, ICML}

\vskip 0.3in
]
\printAffiliationsAndNotice{}
%\title{Inferring the Long-Term Causal Effects of Long-Term Treatments from Short-Term Experiments}
%\author{Allen Tran}
%\affiliation{\institution{Netflix Inc.}\city{Los Angeles}\country{USA}}
%\email{allent@netflix.com}
%\author{}
%\affiliation{\institution{Netflix Inc.}\city{Los Gatos}\country{USA}}
%\email{abibaut@netflixcontractors.com}
%\author{Nathan Kallus}
%\affiliation{\institution{Netflix Inc. \& Cornell University}\city{New York, NY}\country{USA}}
%\email{nkallus@netflix.com}

\date{\today} 

\normalsize{}

\begin{abstract}
We study inference on the long-term causal effect of a continual exposure to a novel intervention, which we term a long-term treatment, based on an experiment involving only short-term observations. Key examples include the long-term health effects of regularly-taken medicine or of environmental hazards and the long-term effects on users of changes to an online platform. This stands in contrast to short-term treatments or ``shocks," whose long-term effect can reasonably be mediated by short-term observations, enabling the use of surrogate methods. Long-term treatments by definition have direct effects on long-term outcomes via continual exposure, so surrogacy conditions cannot reasonably hold. We connect the problem with offline reinforcement learning, leveraging doubly-robust estimators to estimate long-term causal effects for long-term treatments and construct confidence intervals. 
% We present a novel method to estimate the long-term effects of long-term interventions, a critical but challenging task in the absence of long-term randomized control trials. Surrogate methods often rely on shorter-term surrogates and observational datasets, limiting their applicability to short-term interventions. In contrast, 
%Our approach instead learns long-term temporal dynamics directly from short-term experimental data, assuming that the initial dynamics observed persist but avoiding the need for both surrogacy assumptions and auxiliary data with long-term observations. 
%The validity of existing methods for estimating long term effects from short term interventions depend on the sufficiency of surrogates and/or comparability between the short term experimental and longer term observational datasets. However, both assumptions are likely to fail when interventions are permanent, with effects occurring throughout the lifecycle. Instead, we show how to sidestep surrogacy and comparability assumptions with methods that model the underlying dynamics that generate long term outcomes on the experimental dataset alone. We develop an estimator that is root-n consistent and aymptotically normal with weak conditions on nuisance functions.
\end{abstract}
%\vfill


\section{Introduction}
Long-term effects of interventions are often of primary importance yet their direct measurement is hampered by the difficulty of performing long-term randomized control trials. 
% For example, ethical considerations prevent experiments assessing the effects of long term exposure to toxic substances such as mercury. Instead, researchers are often forced to stitch together evidence from observational and epidemiological studies. 
For example, both medical and policy trials are often interested in long-term health or welfare impact, but following subjects for prolonged periods is difficult.
Similarly, businesses in digital settings, constrained by operational considerations and motivated by fast-paced innovation, often use short-run A/B tests to inform decisions that ultimately aim to improve long-term outcomes.

%For example, short term A/B tests are commonly used in digital settings to select among a range of algorithms. Ideally the algorithm with the best long term outcomes would be chosen but budget constraints and the bias towards action lead to shorter term tests.

Surrogate methods offer a route to connect short term tests to their longer term outcomes \cite{athey_index,prentice1989}. These methods rely on the existence of intermediate-term surrogate variables and/or an observational dataset that associates surrogate variables with their eventual long-term outcomes. The key requirements are that the surrogate(s) fully mediate the effect of the treatment on the outcome of interest and that we can identify the effect of the surrogates.

However, the treatment of interest may be explicitly \textit{long-term}, that is, involving a continuous exposure to a novel intervention that extends beyond the length of the experiment. For example, persistent environmental hazards, regular medication, or a change to the user experience in a digital setting. This stands in contrast to short-term treatments, such as a training course or a pharmacological regimen confined in time, whose consequences could reasonably be captured within a short time frame. For long-term treatments, unless the experiment itself (or the measurement of the surrogates) is long-term, surrogate methods are incapable of reliably capturing their effect.

%Because surrogates are by definition shorter term and the link between surrogates and long term outcomes is inferred from a separate dataset under null treatment, the method is suitable only when the underlying intervention itself is \textit{short-term}. 

In this paper, we develop a method that is capable of estimating the long-term effects of long-term treatments from short-term experiments, provided the short-term observations sufficiently characterize the long-term trajectory, even if they do not mediate the effect on it. The method learns long-term temporal dynamics directly from the short-run experimental dataset, which eliminates the need both for the surrogate assumption and for an observational dataset linking surrogates to long-term outcomes. Provided these dynamics persist, this enables the estimation of long-term effects of arbitrary-length treatments, both short and long. In contrast, we show that surrogate methods, even when their assumptions hold, implicitly estimate a truncated effect in our setting, that of a treatment that persists up to the point that surrogates are measured.

In place of the two the key assumptions of surrogate methods (perfect mediation and identification of mediated effect), we make two novel assumptions which connect the problem of estimating long-term effects from experiments with offline reinforcement learning (ORL), which broadly considers the problem of evaluating  ``policies'' on their expected cumulative reward, with evaluation policies potentially differing from the policy generating the data. We make use of the connection with ORL by leveraging recent literature that develops efficient doubly-robust estimators for off-policy evaluation. In particular, we show how long-term causal effects can be estimated from the outcomes of two types of policies: a null treatment policy and a set of policies indexed by $T$, where $T$ denotes the duration of treatment.

The paper is organized as follows. The next section sets up the methodology and provides conditions for identification. Section \ref{sec:estimation} introduces the estimator and conditions for root-$N$ consistency and asymptotic normality. Section \ref{sec:experiments} uses simulated data to evaluate our method against a range of alternatives, as well as exploring robustness to real world complications.  We conclude in Section \ref{sec-conclusion}.

\subsection{Related Literature}\label{sec-litreview}
There exist a long history in Biostatistics of using the response of short-term proxy variables to interventions to infer longer-term effects on a primary outcome of interest. These short-term proxies are referred to as surrogate endpoints and their validity relies on various surrogacy assumptions that share the requirement that the surrogate mediates the treatment effect \cite{prentice1989, vanderweele}.

%``surrogate criterion'': that surrogate variables completely mediate the effect of the intervention on the longer term primary outcome.

However, surrogate assumptions are unlikely to hold for a single surrogate and can potentially lead to sign-reversing bias \cite{chen2007}. The surrogate index literature extends the surrogate method to allow for multiple surrogate variables and the use of observational datasets to infer the relationship between short term surrogates and longer term outcomes \cite{athey_index}. Further extensions to this line of work include learning optimal policies \cite{yang2023targeting}, and combining long and short-term data to tackle confounding \cite{imbens2023longterm,athey2020combining} and improve efficiency \cite{kallus2020role}.

The most related extension similarly focuses on inferring the effects of long-term treatments with short-term experimental measurements \cite{huang-ltt2023}. The approach uses a discrete-time sequential environment with an underlying surrogate-space. To overcome the curse of dimensionality, \cite{huang-ltt2023} assume linearity in both surrogate transitions and surrogate-reward mappings.

The dynamic treatment effects literature similarly seeks to estimate effects for a sequence of treatments \cite{murphy2003, lewis2021, chernozhukov2023automatic}. However, the key difference in our setting is that we aim to estimate treatment effects that extrapolate beyond the horizon of the observed short experiment, whereas dynamic treatment effects methods estimate effects for a horizon that matches the observed data. A related literature exists which aims to undo confounding in a dynamic setting \cite{battocchi,bica2020}. Here, confounding is not a concern since our data come from an experiment where treatment is at worst randomly assigned conditional on the initial state (see Assumption \ref{ass:unconfoundedness}).

We lean heavily on the reinforcement learning literature, which estimates long-term outcomes from the perspective of quantifying the value of different ``policies'' \cite{sutton1998reinforcement}. We make direct use of an estimator from \cite{kallus_uehara} that combines two functions: the $Q$ function, which has a long history in reinforcement learning and the density ratio function \cite{liuNEURIPS2018, ueharaminimax}. 

%Our contribution is to connect the problem of estimating long term effects from interventions to the more general problem of off-policy evaluation in offline reinforcement learning. We reframe the task of evaluation of policies governing the choice of arbitrary actions, to evaluating policies governing treatment or control. Similarly, we implement non-stationary treatment policies of arbitrary duration with a simple stationary treatment policy to mimic long-term experiments. This enables the pairing of powerful ML-based function approximators with efficient and robust estimators, whose asymptotic behavior we study and re-purpose for estimating long term \textit{causal} effects.

%Our contribution is to map the problem of estimating long term effects from interventions to estimating the difference in outcomes from two specific types of policies. Moreover, we do so by avoiding the use of the surrogacy assumptions which allows us to generalize the estimation of long term effects to treatment regimes of any duration.

%The estimator that we use comes directly from \citet{kallus_uehara} who use semi-parametric methods to develop doubly robust estimators to evaluate policies in reinforcement learning. We make direct use of the estimator by mapping the problem of estimating long term average treatment effects from an intervention to estimating the difference in outcomes from two specific types of policies.

\section{Methodology}\label{sec-methodology}
%We're interested in estimating long term effects from an intervention of some duration, where experimental evidence is only available for a shorter duration. %For example, estimating the effect on customer lifetime value from a ``permanent'' change in a recommendation algorithm with evidence from a short-run experiment. 

Let $Y$ denote the long-term outcome of interest and define a treatment policy, $\pi^T$, as a sequence of treatments for $T$ periods and null treatment thereafter.\footnote{The more general case of non-contiguous treatment policies easily fits within our framework, with the addition of more complex notation and a less elegant mapping to stationary state-independent treatment policies (see Section \ref{sec:stat_pols}).} For example, the control policy is $\pi^0$ and a permanent treatment policy is $\pi^{\infty}$. The potential long-term outcome associated with a particular treatment policy, $\pi$, is denoted as $Y(\pi)$.

Our estimand of interest is the average treatment effect of a particular treatment policy: the expected difference in potential long-term outcomes between a $T$-duration treatment policy and the control policy.
\begin{equation}
    \varphi^T = \mathbb{E}\left[Y(\pi^T) - Y(\pi^0) \right]
    \label{eq:ate}
\end{equation}
We assume that we can decompose long-term outcomes into the discounted sum of per period outcomes, normalized so that $Y$ can be interpreted as the weighted average per period potential outcome, weighted towards the present. Let $\gamma$ denote the discount rate, $Y_t$ the per period outcome and $Y_t(a)$ with $a\in \mathcal{A} = \left\{0,1 \right\}$ the per period potential outcome.
\begin{assumption}[Additive rewards]
\begin{equation}
    Y(\pi^T) \equiv (1- \gamma) \sum_{t=0}^\infty \gamma^t Y_t(\mathds{1}_{t < T}) 
    \label{eq:summable_ate}
\end{equation}
\end{assumption}

\begin{figure}
\centering
\begin{tikzpicture}
  \node[circle, draw] (v0) at (-1.5, 3) {$A_0$};
  \node[circle, draw] (v1) at (-3.0, 1) {$S_0$};
  \node[circle, draw] (v2) at (-0.0, 1) {$S_1$};
  \node[circle, draw] (v3) at (-1.5, 1) {$Y_0$};

  \draw [->] (v1) edge (v0)
        (v0) edge (v2)
        (v0) edge (v3)
        (v1) edge[out=-45, in=-135] (v2)
        (v1) edge (v3);
\end{tikzpicture}
\caption{DAG of Observed Experiment}
\label{fig:DAG}
\end{figure}
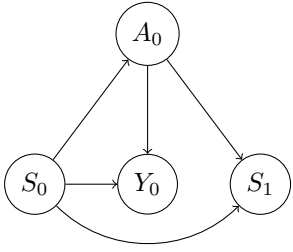

The experiment that generates our data is described in Figure \ref{fig:DAG}. There exists an initial distribution of ``states'', $p_b$, from which initial states, $S_0$, are drawn and in turn which treatment, $A_0$ is assigned. We observe an outcome for the first period, $Y_0$, which depends on both the initial state and treatment assignment. Finally, we observe a transition to a subsequent state, $S_1$, which similarly depends both on the initial state and treatment assignment.

\begin{figure}
\centering
\begin{tikzpicture}
  \node[circle, draw] (v0) at (-2.0, 3) {$A_0$};
  \node[circle, draw] (v4) at (1.0, 3) {$A_1$};
  \node[circle, draw] (v1) at (-3.5, 1) {$S_0$};
  \node[circle, draw] (v2) at (-0.5, 1) {$S_1$};
  \node[circle, draw] (v3) at (-2.0, 1) {$Y_0$};
  \node[circle, draw] (v5) at (1.0, 1) {$Y_1$};
  \node[circle, draw] (s2) at (2.5, 1) {$S_2$};

  \draw [->] (v0) edge (v3)
        (v0) edge (v2)
        (v2) edge (v5)
        (v4) edge (v5)
        (v4) edge (s2)
        (v1) edge[out=-45, in=-135] (v2)
        (v2) edge[out=-45, in=-135] (s2)
        (v1) edge (v3);
\end{tikzpicture}
\caption{DAG of Treatment Policy of Interest}
\label{fig:DAG_policy}
\end{figure}
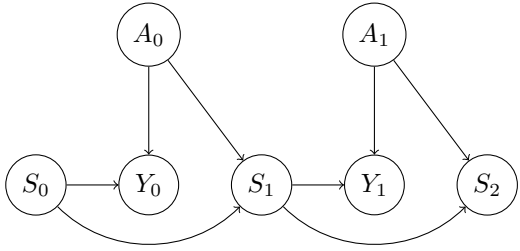

We want to evaluate the ATE with a different treatment policy and potentially a different distribution of initial states, $p_e$, than the experimental distribution. Figure \ref{fig:DAG_policy} depicts the treatment policy and outcomes that we are interested in estimating. Figure \ref{fig:DAG} depicts the short experiment we observe, where treatment is assigned potentially depending on the initial state. In contrast, Figure \ref{fig:DAG_policy} depicts the target policy we want to evaluate where treatment is assigned according to our target policy.

The following two assumptions are assumptions on the experimental design. They are standard assumptions in the causal inference literature and allow us to ``fill in'' the missing counterfactual outcomes with observed outcomes.

\begin{assumption}[Unconfoundedness]
\label{ass:unconfoundedness}
$$\left(Y_t(0), Y_t(1) \right) \indep A_0 \mid S_0$$
\end{assumption}
Given a set of initial states $S_0$, we assume treatment assignment in the experiment is independent of potential outcomes conditional on the initial state, which should be satisfied with experimental data. 
\begin{assumption}[Overlap]
$$\forall s \in \mathcal{S}, a \in \mathcal{A} \colon\; 0 < p_b(s, a) < 1 $$
\end{assumption}
Note that our overlap condition is stronger than in traditional causal inference settings since it technically applies to the entire state-space and not just the initial states. For instance, it requires that treatment is rolled out to all types of users as opposed to being rolled out only to new users. 

%A common scenario when this is satistified is when ``calendar'' time is not part of the covariate space and instead, covariates are either static (e.g levels of engagement with an online service) or refer to relative age (e.g months of membership). Under that scenario, it is relatively easy for a experiment to span the entire covariate space in the first period if the intervention applies to all units.

The next two assumptions depart from existing methods and allow us to extrapolate beyond the short term, using only data from the experiment. The states are assumed to satisfy the Markov property. The Markov assumption is implicitly a requirement that the state-space is sufficiently rich.
\begin{assumption}[Markov property in states and actions] 
% For all $$, we have
    \begin{align*}
    &\forall s \in \mathcal{S}^t, a \in \mathcal{A}^{t-1}:\\
    &\qquad p\left(s_t| s_{t-1}, a_{t-1}, \ldots s_0, a_0 \right) = p\left(s_t | s_{t-1}, a_{t-1} \right)
    \end{align*}
\end{assumption}

Define $p(y, s|\pi^T;t)$ as the marginal distribution of the states $s$ and outcomes $y$ ``induced'' by projecting the transition probabilities $t$ periods from the initial distribution of states, $p_0(s_0)$, under the policy $\pi^T$.
\begin{align}
    p(y, s|\pi^T;t) &= \int_{s_0, \ldots, s_{t-1}} p_0(s_0) \prod_{k=1}^{t} p(s_k|s_{k-1}, \mathds{1}_{k-1 < T}) \nonumber \\
    &\quad \times p(y|s, \mathds{1}_{t < T})
\end{align}
%\begin{multline}
%    p(y, s|\pi^T;t) = \\ \int_{s_0, \ldots, s_{t-1}} p_0(s_0) p(s_1|s_0, \mathds{1}_{0 < T}) \ldots  p(s|s_{t-1}, \mathds{1}_{t-1 < T}) p(y|s, \mathds{1}_{t < T})
%\end{multline}
\begin{assumption}[Stationarity]
    \begin{multline*}
    \forall t, y, s, \pi^T:\\ 
    p_t(y, s|\pi^T) - p_t(y, s|\pi^0)  = p(y, s|\pi^T; t) - p(y, s|\pi^0; t)
    \end{multline*}
\end{assumption}
Stationarity assumes that the \textit{difference} in the marginal distributions of $s$ and $y$ with respect to any treatment policy $\pi^T$ and the control policy $\pi^0$ at each period match those induced by the Markov transition probabilities. The assumption allows \textit{levels} of these distributions to change, as long as the changes apply equally to treatment and control populations.

%Combined with overlap, this implies that estimands that depend on the difference in $t$-dated conditional probabilities can be calculated from our experimental data alone. At heart, stationarity is similar to the comparability assumption in the surrogate literature. Although both assumptions ensure that historical data are helpful for extrapolating into the future, stationarity is a weaker than comparability in the range of data that is required to be relevant.

%Comparability requires the existence of a 1) long term observational dataset that has 2) the same conditional probabilities as the experimental dataset. Because we're interested in long term outcomes, this implicitly means that the observational dataset is as old as the duration of the long term effect.\footnote{An alternative is a shorter duration dataset that extrapolates into the future by making additional assumptions, e.g a parametric survival model.} Hence by definition, the observational dataset is much older than the experimental dataset and less likely to be relevant. 

%Regardless, note that we only require stationarity in the difference between treatment and control and not on non-differenced conditional probabilities. Hence the method is robust to changes in the environment, except those that differentially affect treatment or control.

\subsection{Identification}
\label{sec:ident}
A necessary step in estimating the average treatment effect of a treatment policy depicted in Figure \ref{fig:DAG_policy} is to express the estimand as a function of observable data available from an experiment. In particular, we assume the observable data consists of $N$ i.i.d. tuples ($a, s, y, s')$ generated from the process illustrated in Figure \ref{fig:DAG}.

To do so, we will exploit the fact that the environment and assumptions above describe a Markov Decision Problem (MDP). Our setup is an MDP with a binary action space, $\mathcal{A}$, state-space $\mathcal{S}$, expected reward emission function, $p(y|s, a)$, state-transition kernel $p(s'|s, a)$ and a \textit{non-stationary} policy, $\pi^T_t: \mathcal{A} \times \mathcal{S}  \times \mathbb{N}^+ \rightarrow [0, 1]$.

Leaning on the framing as a MDP, we can summarize the cumulative discounted outcomes recursively using the state-action value function (the $Q$ function) defined as follows.
\begin{equation}
    q^{^T}_t(s, a) \equiv \mathbb{E}_y\left[y|s, a\right] + \gamma \mathbb{E}_{s'\sim p(\cdot|s, a)}\left[q^T_{t+1}(s', \mathds{1}_{t+1 < T}) \right]
    \label{eq:q_nonstat}
\end{equation}
The superscript of the $Q$ function denotes the associated policy is the $T$-duration treatment policy $\pi^T$ and the subscript denotes the dependence on time. The $Q$ function as described in \eqref{eq:q_nonstat} is non-stationary because the $T$-duration treatment policy is non-stationary. Note that for fixed policies, either persistent treatment or control, the $Q$ function is stationary since the underlying policy is constant over time.

\begin{theorem}[Identification by $Q$ function with non-stationary policy]\label{thm:nonstat_ident}
Suppose Assumptions 1-4 hold. Then the average treatment effect of a $T$-duration treatment policy is composed of the following function of observable data.
\begin{equation}
    \varphi^T = (1-\gamma) \mathbb{E}_{s \sim p_0(\cdot) }\left[q_0^T(s, \mathds{1}_{0 < T}) - q^0(s, 0)\right]  \\
    \label{eq:pot_outcome_with_nonstatq}
\end{equation}
\end{theorem}
Theorem~\ref{thm:nonstat_ident} transforms the complex task of estimating infinite horizon per-period outcomes into a straightforward computation of the $Q$ function, weighted appropriately and evaluated at the initial state and action. The proof is provided in the Appendix in Section \ref{sec:proof_ident}.

%The key insight is that if we observe state transitions for all pairs of states for each of treatment and control (the actions), then we can identify the probability of a unit being in a state at any time period. Since we also observe the expected outcome conditional on the state, we can characterize the distribution of expected outcomes across time. Stationarity guarantees that these conditional probabilities are constant over time (up to differencing) and hence we can estimate these distributions on our short-term experimental data. 

%As an illustrative example, consider estimating the effect of a permanent intervention policy against the control policy. These policies are simpler to evaluate since these policies are stationary and are not state-dependent. By assumption, the experiment allocates both treatment and control for each state. Hence we can simply estimate state \textit{value} functions separately for each of the treatment and control arms and difference the two making sure to account for any initial imbalance in the treatment assignment.

\subsubsection{Stationary Policies}\label{sec:stat_pols}

Theorem~\ref{thm:nonstat_ident} is challenging to use directly since the $Q$ function in Equation \eqref{eq:pot_outcome_with_nonstatq} is difficult to estimate due to it inheriting non-stationarity from the underlying $T$-duration treatment policy.\footnote{For our specific form of non-stationarity, one would need to first estimate the $Q$ function at $T$ under the deterministic control policy, then the $T-1, \ldots, 0$ $Q$ functions in order under the deterministic treatment policy.} Instead, we prove the existence of an equivalent stationary stochastic policy and construct a computationally efficient approximation. With such a stationary policy, we construct a practical version of Theorem~\ref{thm:nonstat_eq_stat} in Corollary \ref{cor:stat_id} which uses a stationary policy making the $Q$ function tractable.

A key concept we need is that of an occupancy measure, the discounted fraction of time an agent spends in state $s$ and action $a$.
\begin{equation}
    \rho_{\pi, \gamma}(s, a) \equiv (1-\gamma) \sum_{t=0}^\infty \gamma^t \rho_{\pi, t}(s, a)
    \label{eq:occ_sa}
\end{equation}
Similarly, there exists the state occupancy measure which is Equation \eqref{eq:occ_sa} marginalizing out actions. The occupancy measures provide a way to express the cumulative discounted outcomes, a summation across time, instead as a single point-in-time weighted average of outcomes.

\begin{equation}
    (1-\gamma)\mathbb{E}_{p_0, \pi }\left[\sum_{t=0}^\infty \gamma^t y_t \right] = \mathbb{E}_{s, a \sim \rho_{\pi, \gamma}(\cdot), y \sim p(\cdot|s, a)}\left[ y \right]
\label{eq:equal_occ_equal_cumsum}
\end{equation}
An implication of Equation \eqref{eq:equal_occ_equal_cumsum} is that two policies that lead to the same occupancy measures will have the same cumulative discounted rewards.

\begin{lemma}[Stationary equivalents of non-stationary policies]\label{thm:nonstat_eq_stat}
For any non-stationary policy $\pi = \pi_0, \pi_1, \ldots$, there exists a stationary policy $\bar{\pi}$ that generates the same occupancy measure. In particular construct a stationary policy as follows:
\begin{equation}
    \bar{\pi}(a|s) = \frac{\rho_{\pi, \gamma}(s, a)}{\rho_{\pi, \gamma}(s)},
    \label{eq:stat_pol}
\end{equation}
then
\begin{equation}
    \rho_{\pi, \gamma} = \rho_{\bar{\pi}, \gamma}.
\end{equation}
\end{lemma}
See \cite{bertsekas-dynamicProgramming} for a proof.

\begin{theorem}[Stationary T-Duration Treatments]
For a non-stationary policy $\pi^T$ that sets $a=1$ for $T$ periods and $a=0$ thereafter, ($i$) there exists an equivalent stationary stochastic policy $\bar{\pi}^T$ that yields the same cumulative discounted reward and ($ii$) the average of that stationary stochastic policy across states is $1 - \gamma^T$. 
\end{theorem}
\begin{proof}
    Let $\pi^T$ be an arbitrary non-stationary policy. That non-stationary policy leads to associated occupancy measures, $\rho_{\pi^T, \gamma}$. Construct a candidate stationary policy:
    \begin{equation}
        \bar{\pi}^T(s,a) = \frac{\rho_{\pi^T, \gamma}(s,a)}{\rho_{\pi^T, \gamma}(s)}
    \end{equation}
    Lemma \ref{thm:nonstat_eq_stat} shows that $\bar{\pi}^T$ leads to an equivalent occupancy measure, and hence will result in the same expected cumulative discounted reward as under $\pi^T$.

    For ($ii$), the weighted average treatment policy 
    across states is:\phantom\qedhere
    \begin{align*}
        \textstyle\int_s \bar{\pi}^T(a|s) \rho_{\bar{\pi}^T, \gamma}(s) ds &\textstyle= \int_s  \rho_{{\pi}^T, \gamma}(s, a) ds \\
        &\textstyle=\int_s (1-\gamma) \sum_{t=0}^\infty \gamma^t \rho_{{\pi}^T, t}(s, a) ds \\
        &\textstyle=(1-\gamma) \sum_{t=0}^\infty \gamma^t \int_s \rho_{{\pi}^T, t}(s, a) ds \\
        &\textstyle=(1-\gamma) \sum_{t=0}^T \gamma^t  =1 - \gamma^T.\qed
    \end{align*}
\end{proof}
Intuitively, a stationary policy with a constant treatment probability of 0 corresponds to a control policy indexed by $T=0$. As $T$ increases, so does this probability, and as $T \rightarrow \infty$, it approaches 1. In general, constructing the exact state-dependent equivalent stationary policy is intractable since it requires estimating the occupancy measures under the $T$-duration treatment policy. Instead, we suggest using the state-independent policy, $\forall s\colon\;\bar{\pi}^T(a|s) = 1 - \gamma^T$, which offers a practical and computationally efficient approximation.

Constructing a stationary policy from a $T$-duration policy via Equation \eqref{eq:stat_pol} leads to a stationary $Q$ function, equivalent in occupancy measures and expected outcomes to the non-stationary $Q$ function in Equation \eqref{eq:q_nonstat}, when starting from the same initial distribution of states.
\begin{equation}
    q^{^T}(s, a) \equiv \mathbb{E}_y\left[y|s, a\right] + \gamma \mathbb{E}_{s'\sim p(\cdot|s, a), a' \sim \bar{\pi}^T(\cdot|s')}\left[q^T(s', a') \right]
    \label{eq:q_stat}
\end{equation}
Hence we can state a stationary version of Theorem~\ref{thm:nonstat_ident}, with a stationary and hence learnable $Q$ function.

\begin{corollary}[Identification by Stationary-policy Q]
Suppose Assumptions 1-4 hold. Then the expected average treatment effect of a $T$-duration treatment policy is equal to expectation over the difference of $Q$ functions, associated with the equivalent stationary policy, $\bar{\pi}^T$ and the control policy.
\begin{equation}
    \varphi^T = (1-\gamma) \mathbb{E}_{s \sim p_0(\cdot), a \sim \bar{\pi}^T(\cdot|s)}\left[q^T(s, a) - q^0(s, 0) \right]  
    \label{eq:ate_with_q}
\end{equation}
\label{cor:stat_id}
\end{corollary}
\subsection{Comparison to Surrogate Index Method}\label{sec:surrogate_comp}
Within the current environment, it is instructive to pinpoint the key difference between our method and the surrogate index method \cite{athey_index}. Assume the setup above where we observe everything up to some period $t$, where we observe only the transition to the $t$th period state. In other words, we observe $N$ tuples of $(s_0, y_0, a_0, \ldots, s_t)$.

We focus on the difference in outcomes for a permanent treatment policy for periods past $t$ since the covariate adjusted difference in means will recover the treatment effect prior to $t$. Following the assumptions in Section \ref{sec:ident}, the true potential outcome for period $t+k$ where $k > 0$, can be expressed as
\begin{align}
 \mathbb{E}\left[Y_{t+k}(1) \right] &= \int_{s_t, s, y} y p(y \mid s, a=1) \nonumber \\
 &\quad \times p(s \mid s_t, a=1;k) p_t(s_t \mid a=1).
 \label{eq:decomp_ate}
\end{align}
where $p(s'|s, a;k)$ is the transition kernel projected $k$ periods ahead starting from $s$.

A surrogate method that relies on an observational dataset will instead calculate the expectation of the $t+k$ period outcome conditional on the distribution of $s_t$ from the experiment but also in part on a probability model, $p^o$ learned from an observational dataset.
\begin{align}
    &\mathbb{E}\left[Y_{t+k}(1)]\right] \leftarrow \mathbb{E}_{s_t}\left[\mathbb{E}_Y^o\left[Y_{t+k} |s_t, a=0\right]|a=1\right]  \nonumber\\
    & = \int_{s_t, y_{t+k}} y_{t+k} p^o(y_{t+k} \mid s_t, a=0)p_t(s_t| a=1)  \nonumber\\
    %\mathbb{E}\left[y_1]\right] &\leftarrow \sum_{y_1} y_1 p^o(y_1 \mid s_0, a=0)p_0(s_0)  \nonumber\\
    &= \int_{s_t, s_{t+k},y_{t+k}} y_{t+k} p^o(y_{t+k} \mid s_{t+k}, a=0) \nonumber \\
    & \quad \times p^o(s_{t+k}|s_t, a=0) p_t(s_t| a=1)
    \label{eq:obs_estimate}
\end{align}
Note that when the observational model is used, it conditions on null treatment since our novel treatment doesn't exist in the observational dataset. The comparability assumption ensures that the observational and experimental probabilities are equal, $p^{o} = p$. 

Equation \eqref{eq:obs_estimate} makes it clear that the surrogate estimate only captures the partial treatment effect that is mediated through the surrogate, $s_t$. For periods beyond the measurement period of the surrogate, it misses that \textit{permanent} interventions may alter (i) state transitions and hence affect the distribution of future states, $p(s_{t+k}|s_t, a=1) \neq p(s_{t+k}|s_t, a=0)$ and (ii) the contemporaneous relationship between state and outcome, $p(y|s, a=1) \neq p(y|s, a=0)$.\footnote{Of course, these effects diminish as the period of surrogate measurement increases. But this point is moot as the problem at hand is to estimate long-term effects on short term experimental measurements.}

Hence surrogate index methods capture long-term effects, but only for treatment durations up to the time when the surrogate is measured. The effects captured are indirect long-term effects due to the persistence of initial treatment effects. We verify this experimentally in Section \ref{sec:experiments}.

%This is particularly troublesome for intervention policies lasting beyond the duration of the experiment, which span a wide range of real-world use cases.

\section{Estimation}
\label{sec:estimation}
We want to estimate the long-term average treatment effect via the $Q$ function in Equation \eqref{eq:ate_with_q}. With discrete states, we can solve for $Q$ exactly via dynamic programming methods subject to computational constraints \cite{sutton1998reinforcement}. But when the state-space is large or continuous, we need to rely on machine learning techniques to approximate the $Q$ function.

It is well known that relying on ML-based estimators in a statistical estimand may lead to bias due to overfitting and regularization techniques used in training \cite{cherno_dml}. Hence we develop a double ML based estimator centered around the efficient influence function \cite{kallus_uehara}. The estimator is $N^{-\frac{1}{2}}$ consistent and doubly robust with respect to ML-learned $Q$ and density ratio functions, which are only required to converge at slow rates.

\subsection{Efficient Influence Function Based Estimator}
The estimator we propose is the naive plug-in estimator with a bias correction term based on the efficient influence function. The efficient influence function for one half of the estimand (the potential outcome under the policy $\pi$) is a function of the observed tuple $\left( s, a, y, s' \right)$,  a stationary policy $\pi$ and the nuisance functions $q$ and $w$, representing the $Q$ and density ratio functions.
\begin{multline}
    \phi^\pi(s, a, y, s'; q, w) = -\varphi^{\pi} + (1-\gamma) \mathbb{E}_{s \sim p_0(\cdot)}\left[q^\pi(s, a) \right] \\ 
    +\frac{{\pi}(a|s)}{p_b(a|s)} w(s) \left(y + \gamma \sum_{a' \in \mathcal{A}}{\pi}(a'|s') q^\pi(s', a') - q^\pi(s, a) \right)
    \label{eq:eif}
\end{multline}
The efficient influence function for the estimand is simply the difference of the respective efficient influence functions for treatment and control.
\begin{multline}
    \phi(s, a, y, s';q, w, \pi, \pi^0) = \phi^\pi(s, a, y, s';q, w) \\
    - \phi^{\pi^0}(s, a, y, s';q, w)
\end{multline}
The density ratio function is defined as
\begin{equation}
    w(s) \equiv \frac{\rho_{\pi, \gamma }(s)}{p_b(s)}.
    \label{eq:def_w}
\end{equation}
An intuitive description of the density ratio function is the occupancy measure under the policy relative to the probability density function under the experiment.

The bias-corrected estimator is
\begin{multline}
    \varphi^{\pi}_{BC} \equiv (1-\gamma)\mathbb{E}_{s \sim p_0(\cdot), a \sim \pi(\cdot|s)}\left[q^\pi(s, a) -q^{\pi^0}(s, 0) \right] \\
    + \mathbb{E}\left[ \phi(s, a, y, s';q, w, \pi, \pi^0) \right]
    \label{eq:plug_in_estimator}
\end{multline}
where the final term is a bias correction term that undoes any asymptotic bias from using ML-based estimators of the nuisance and value functions \cite{kennedy2023semiparametric}.

\subsection{Asymptotic Properties}
By design, we specified the ATE as a function of a well studied objective in reinforcement learning: the normalized discounted outcomes associated with a policy. Hence we can make use of results from \cite{kallus_uehara} who propose efficient, doubly robust estimators for off-policy evaluation using semiparametric methods. This section summarizes the relevant results.

The difference between the bias-corrected estimator and the true estimand can be decomposed into three components: a central limit theorem term, an empirical process term and a second order remainder term. The key is to show that the empirical process term and the second order remainder term converge to zero faster than the central limit theorem term. 

The empirical process term is $o_p(N^{-\frac{1}{2}})$ if we use cross fitting in estimation.  Cross fitting involves splitting the data into $K$ partitions, estimating nuisance functions on the $K-1$ held out partitions, evaluating the estimator for each single partition and finally averaging over the $K$ estimators to get the final estimate. 

For brevity, we refer to \cite{kallus_uehara} for details on controlling the second order remainder term.

\begin{theorem}[Double Robustness] If either one of $\|\hat{q} - q \|_2 = o_p(1)$ or $\|\hat{w} - w \|_2 = o_p(1)$ holds, then $\hat{\varphi}_{BC} - \varphi = o_p(1)$.
\end{theorem}

Double robustness implies that we only need to ``correctly'' estimate one of either the $Q$ or the density ratio functions to ensure our bias corrected estimator is consistent. %Intuitively, if the estimate of the $Q$ function is correct, $\hat{q} = q$, then $R = 0$ since simple algrebra shows that $P \hat{\phi}^\pi(s, y, s', x, a';q, \hat{w})  = -\hat{\varphi}^\pi + \varphi^\pi$. On the other hand, when only the density ratio functions are correct, Lemma 1 (a characterization of $w(s)$) is required to show that $R=0$ when $\hat{w} = w$.

%Denoting $\eta(s, a)$ as the ratio of the policy of interest and policy governing the experiment, we have the following property of the density ratio function.
%\begin{lemma}[Characterization of $w$]\label{lem:ratio-estimation}
%Define 
%\begin{multline}\label{eq:estimating}
 %   L(w',f)=\mathbb{E}_{}[\gamma w'(s)\eta(s,a)f(s')-w'(s)f(s)]\\
%    +(1-\gamma)\mathbb{E}_{s \sim p_0(\cdot)}[f(s)]
%\end{multline} 
%Then, for $w'=w$, we have $L(w',f)=0$ for any $f$. 
%Conversely, if $L(w',f)=0$ for all $\lambda_{\mathcal S}$-square-integrable functions $f$ and there is a unique solution $g$ to the integral equation 
%\begin{align*}
    %0=\gamma \int p(s'|s)g(s)\mathrm{d}\lambda_{\mathcal S}(s)-g(s')+(1-\gamma)p_0(s'),
%\end{align*}
%then, $w'(s)=w(s)$. 
%\end{lemma}

%For each of $w^1$ and $w^0$ and the choices for $f_w$ of $\hat{v}_1$ and $\hat{v}_0$ respectively, by using Lemma 1 we can similarly show that the terms in $R$ cancel such that $R=0$ when only the density ratios are correct.

\begin{theorem}[Asymptotic Normality and Efficiency]
    Suppose that (i) $\hat{q}$ and $\hat{w}$ converge to $q$ and $w$ in probability at rates such that the product of those rates is $o_p(N^{-\frac{1}{2}})$ and (ii) the propensity score $p_0(a|x)$ is known. Then the bias-corrected estimator is asymptotically normal and efficient.
    $$
    \sqrt{N} \left(\hat{\varphi}^\pi_{BC} - \varphi^\pi \right) \xrightarrow{d} \mathscr{N}\left(0, \phi^{2} \right)
    $$
\end{theorem}

Crucially, the convergence rate requirements on the $Q$ and density ratio function estimates are each slower than square-root which enables the use of a range of ML algorithms along with techniques such as regularization. If the propensity score needs to be estimated, then the rate requirement on the density ratio function instead applies to the product of the density ratio function and the propensity score.

\label{sec:2ndorder}
\subsection{Q Function Estimation}\label{sec-MCM}

If states are finite, then dynamic programming techniques can solve for the $Q$ function exactly given the availability of transition probabilities. However, since we potentially have continuous states or a large state space which is subject to the curse of dimensionality, we need to use ML techniques that parameterize the $Q$ function.

An obvious choice is the family of Temporal Difference (TD) algorithms used for policy evaluation. TD algorithms estimate the $Q$ function on a dataset of state transitions, actions and rewards, as available in an experiment. In particular, it requires a dataset of $\left(s_i, a_i, y_i, s_i'\right)$ tuples for $i=1, \ldots, N$ units and parameterizes the $Q$ function with a vector of parameters, $\theta_q$. Hence
$$
q^\pi(s, a; \theta_q) \approx q^\pi(s, a).
$$

Using the definition of the $Q$ function from Equation \eqref{eq:q_stat}, we can form the TD error term
\begin{equation}\label{eq:td_error}
    L_Q(s, a, y, s') = y + \gamma \mathbb{E}_{a' \sim \pi(\cdot|s'}\left[q^\pi(s', a'; \theta_q \right] - q^\pi(s, a;\theta_q)
\end{equation}
whose expectation is zero for the true $Q$ function.

Within the family of TD methods, various approaches have been proposed which center on minimizing the TD error \cite{sutton1998reinforcement}. Framing the TD error in Equation \ref{eq:td_error} as an estimating equation, one can use techniques from M-estimation to derive asymptotic properties such as asymptotic normality and consistency. For example, \cite{kallus_uehara} derive the asymptotic lower bound for an M-estimator that seeks to minimize a weighted form of Equation \eqref{eq:td_error}.

\subsection{Density Ratio Estimation}
The use of density ratio functions in reinforcement learning is relatively new, finding recent use in methods for efficient off-policy evaluation \cite{liuNEURIPS2018, ueharaminimax, kallus_uehara}. Their estimation has centered on the following relationship:
\begin{multline}\label{eq:wf_relationship}
    L_W(f, w) = \\
    \mathbb{E}_{s, a, y, s' \sim p_0, a' \sim \pi(\cdot|s')} \left[w(s,a)\left( \gamma f(s', a') - f(s, a)\right) \right] \\
    - (1-\gamma)\mathbb{E}_{s \sim p_0, a \sim \pi(\cdot|s)}\left[f(s,a) \right],
\end{multline}
which equals zero for the true density ratio function and can be derived from the definition of the $Q$ function.

Under some mild technical conditions, \cite{ueharaminimax} show that if $L_W(f, \hat{w})=0$ for all $f$, then $\hat{w} = w$. Moreover, the reverse also holds, that the true density ratio function is the only function for which the statement is true. 

This leads to a Minimax-style estimator, with two function classes, $\mathcal{F}$ and $\mathcal{W}$, each encompassing the discriminator and the density ratio functions.
\begin{equation}
    \hat{w}(s, a) = \argmin_{w \in \mathcal{W}} \max_{f \in \mathcal{F}} L_W(f, w)^2
\end{equation}
%Loosely speaking, finding a $\hat{w}$ that sets the Minimax objective close to 0 guarantees that $\hat{w} \approx w$ since the inner maximization bounds the error across all $f \in \mathcal{F}$ \cite{ueharaminimax}. 

Minimax estimators can be challenging to implement due to the inner maximization. Fortunately, the Minimax objective can be reduced to avoid the inner maximization in two cases \cite{ueharaminimax}. First, when the function classes of the density ratio functions and the discriminator are linear under the same feature maps for state and actions. Second, when the discriminator function class corresponds to a reproducing kernel Hilbert space.

\section{Experiments}\label{sec:experiments}
To demonstrate the effectiveness of the ORL method, we perform experiments on simulated data.\footnote{Code and data for the experiments is available at: \href{https://github.com/allentran/long-term-ate-orl}{https://github.com/allentran/long-term-ate-orl}
.} First, we empirically validate some of the theoretical results  on a simple MDP which allow us to precisely control aspects of the data generating process so that we can see the effects of gradually relaxing the adherence to our assumptions. Second, we evaluate the method using a simulator of sepsis where effect sizes and randomness reflect real-world settings \cite{pmlr-v97-oberst19a}.

%The results confirm that the surrogate index method recovers the true long-term ATE, albeit only for a treatment of duration of a single period. However, when estimating effects from treatments of duration longer than a single period, the estimated effect is constant and hence its bias grows with the underlying treatment duration. On the other hand, the ORL method estimates match the true ATE across all treatment durations, growing as the duration of treatment extends from a single period to permanent. Furthermore, as expected from theory, (i) the effect of an unobserved hidden state can induce bias that increases with the relevance of the hidden state and (ii) decreased ``coverage'' of states in the experiment lead to greater variance in the estimates.

%In experiments under the more realistic sepsis simulator setting, the ORL method accurately estimates the long-term ATE although performance is degraded at short treatment durations, where the approximation by a non-stationary policy of the true non-stationary T-duration treatment policy is worse.

\subsection{Simple MDP}
\paragraph{Simulation Details}~ We generate data with characteristics that mirror a streaming video on demand service. We construct a single state Markov chain mirroring the dynamics of on-service tenure (state) and subscription revenue (reward). On-service tenure increases if members do not churn, which we model as a drift-diffusion process with positive drift:
\begin{equation}
    s_t = s_{t-1} + \mu + \tau_s a_t + \sigma_s w_t
    \label{eq:sim_states}
\end{equation}
where $w_t \sim \mathcal{N}(0, 1)$ and we restrict $s_t \in [0, 1]$ by clipping.

Since longer tenured members often have higher yet capped per period revenue, we set the reward to be a diminishing scalar multiple of the state.
\begin{equation}
    y_t = \alpha s_t^\theta + \tau_y a_t + \sigma_r e_t
    \label{eq:sim_reward}
\end{equation}
where $e_t \sim \mathcal{N}(0, 1)$.

Treatment effects both increase the probability of a transition to a higher state and increase average per-period rewards conditional on the state. Table \ref{tab:freq} in the Appendix lists the parameter values used in the experiments. 

Within this setting, we (i) benchmark a variety of estimators across a range of underlying treatment durations, (ii) determine the effect of a hidden state on estimates while varying the ``relevance'' of the hidden state and (iii) assess the impact of decreased support in the experimental data for states whose treatment effects we wish to evaluate. 

\paragraph{Treatment duration}~ We simulate trajectories over a long horizon to mimic an infinite horizon where treatments vary in duration from 1, 6, 12, 24, 48 and 120 periods.\footnote{We label the 120 period duration as $\infty$ as the two durations are quantitatively equivalent given our choice of the discount rate.} Note that the estimand is infinite horizon, so that even for short duration treatments, we are interested in the long-term ATE.

In addition to the ORL method, we evaluate a naive method and two variants of the surrogate index method. The naive method assumes the 1-period experimental ATE is constant throughout the duration of treatment and zero thereafter. The first surrogate index variant sees two periods of the experiment, resembling access to data from a short experiment whereas the second variant sees the experiment for the duration treatment is applied.

\begin{figure}[h]
    \centering
    \includegraphics[width=0.5\textwidth]{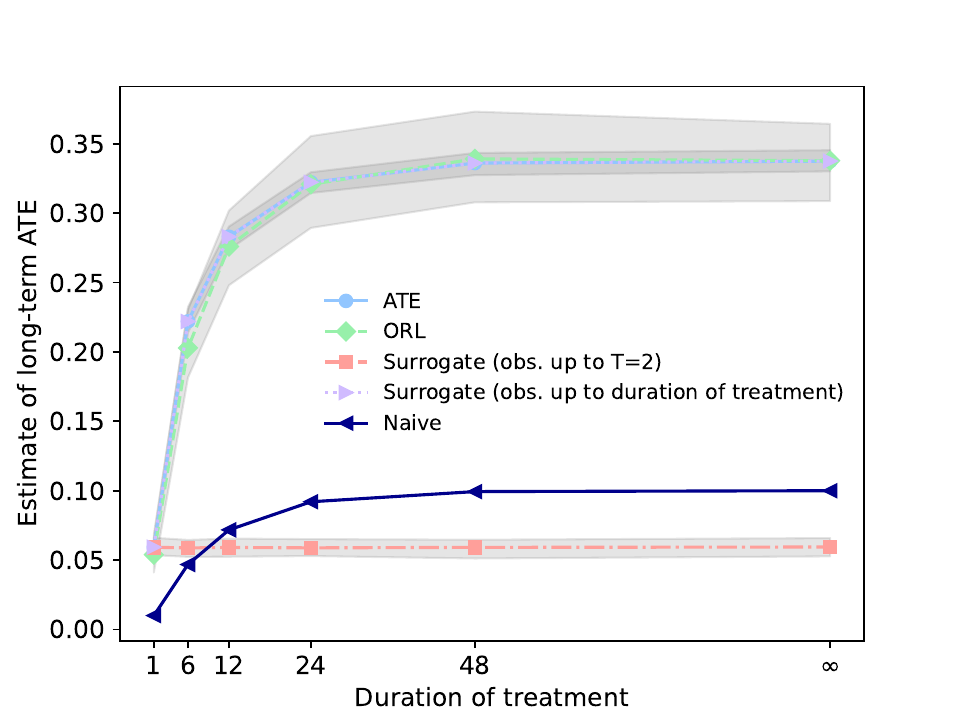} % Replace "filename" with your image file's name, without the extension
    \caption{Estimates of the ATE for differing treatment durations}
    \label{fig:xp-duration}
\end{figure}

Figure \ref{fig:xp-duration} shows estimates across the estimators for varying durations of treatment. Since the treatment increases both the rewards per state and the likelihood of a positive state transition, the true ATE (light blue) increases with the duration of the treatment. Estimates from the ORL method (green) match the true ATE, as the duration of treatment extends all the way to a permanent intervention. 

Conversely, the surrogate index method estimates outcomes as if treatment is applied at most, up to the point at which the last surrogate is measured. Hence the surrogate index method with the last surrogate observed in the second period estimates the ATE for treatment active for a single period. The estimate is constant and therefore underestimates the true ATE for $T>1$. The surrogate method (purple) matches the true ATE but crucially, requires an infeasible duration of measurement, with surrogates measured up to the target duration of treatment.

%While both methods match the true ATE, their requirements on measurement are vastly different. The ORL method can estimate the ATE for an arbitrary duration treatment regime with a short experiment. Its data requirements remain static with respect to the treatment duration of interest.

\paragraph{Lack of coverage in experimental data}~ The ORL method uses data from an experiment with a particular distribution over states and actions to estimate causal effects for a potentially different distribution of initial states. The two distributions can differ vastly in situations such as medical trials where it may be difficult to enrol some segments of the population of interest. This difference is summarized by $w$, the density ratio function in Equation \eqref{eq:def_w} and explodes as the coverage in the experimental distribution goes to zero. Accordingly, the asymptotic variance of the ORL estimator increases with the square of the density ratio function. %Intuitively, if we lack support in our experimental distribution for states we want to evaluate, the variance of the estimator increases. 

\begin{figure}[h]
    \centering
    \includegraphics[width=0.45\textwidth]{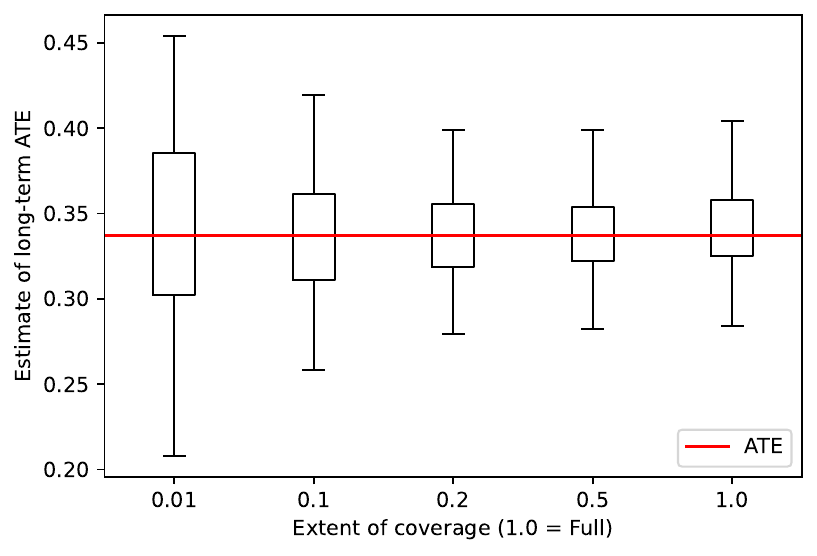} % Replace "filename" with your image file's name, without the extension
    \caption{ORL estimates for differing state coverage in experimental data}
    \label{fig:xp-coverage}
\end{figure}

To assess the effect of a lack of ``coverage'', we downweight observations with states within an interval of 0.2 and 0.8 by various degrees and resample with replacement to keep the number of observations fixed.\footnote{The evaluation state distribution is uniform between 0 and 1.} Results in Figure \ref{fig:xp-coverage} show that as expected, the variance of estimates increases as coverage decreases, although only having an effect when the missing data is downweighted 1:100.

\paragraph{Effects of a hidden state}~ A key assumption is that we observe all the states underlying the MDP. The existence of a hidden state may lead to bias and increased variance in estimates of the ATE since the $Q$ and density ratio functions are underspecified.

The impact of the hidden state depends largely on how independent the hidden state is and whether it affects state transitions that govern reward emissions. To demonstrate this, we add an additional state and alter state transitions via Equation \eqref{eq:states-hidden}. When $\rho=0$, the hidden state is decoupled from the MDP and functions as noise. When $\rho=1$, the state which governs reward emissions depends entirely on the hidden state and hence the hidden state needs to be observed for accurate $Q$ function estimation.

\begin{equation}
    \mathbf{s_t} = P \mathbf{s_{t-1}} + \mu + \tau_s a + \sigma_s \mathbf{w_t}
    \label{eq:states-hidden}
\end{equation}
where bolded variables represent column vectors and 
\begin{equation}
P = \begin{bmatrix} 1 - \rho & \rho \\ 0 & 1 \end{bmatrix}.
\end{equation}
\begin{figure}[h]
    \centering
    \includegraphics[width=0.45\textwidth]{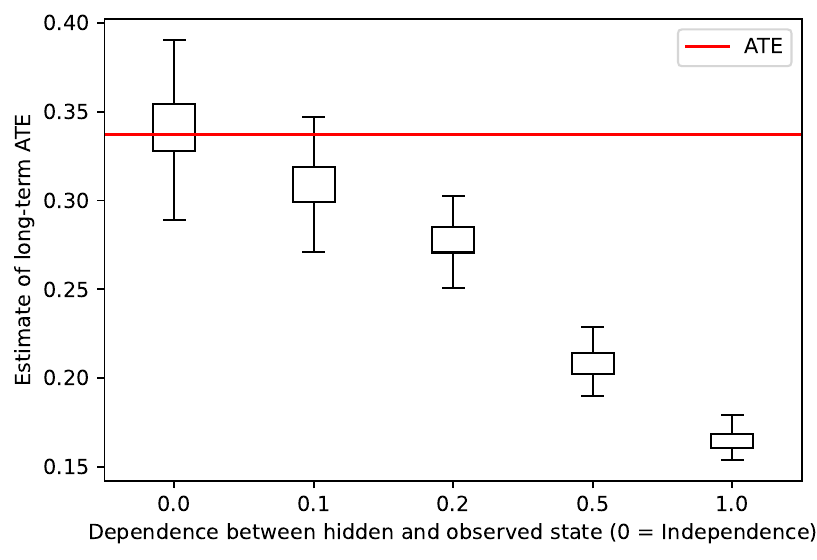} % Replace "filename" with your image file's name, without the extension
    \caption{ORL estimates with unobserved hidden state of varying relevance}
    \label{fig:xp-hidden}
\end{figure}

Figure \ref{fig:xp-hidden} shows that increasing the relevance of a hidden state introduces severe bias. Although not depicted, the cause for the bias is downwards bias in the estimate of rewards from the treatment. As $\rho \rightarrow 1$, the model has difficulty predicting state transitions which leads to predicting transitions closer to the mean of the state distribution, understating treatment effects. This highlights the importance of using a large state-space and methods capable of overcoming the curse of dimensionality.

\subsection{Sepsis Simulator}
To evaluate our method in a more realistic setting, we generate data from a simulator of sepsis \cite{pmlr-v97-oberst19a}. The simulator is based on an underlying MDP with measurements of vital signs and previous treatment as states and rewards of $+1$ and $-1$ respectively upon patient discharge and death. Using a more realistic environment allows us to assess more quantitative dimensions of performance, such as the size of confidence intervals compared to effect sizes. To maintain the constraint of realism, we use feature representations that are human-interpretable (e.g each vital sign and its measurement is a feature) as opposed to using one-hot encoded representations of each discrete state.

%As the sepsis simulator relies on discrete states, with enough data, a one-hot encoded representation of the states identifying each discrete state would recover the ATE exactly. To ensure a realistic setting where the true states are unknown, we keep states in the original human-interpretable format, with each vital sign and its measurement as a dimension of the state-space. In addition, the simulator generates trajectories that are episodic since death and discharge are terminal events whereas our method applies to tasks that are continuing. To handle this, we force outputs from the learned $Q$ function to be zero for terminal states.\footnote{Similarly, we could have introduced another dimension of the state-space to be a boolean terminal state indicator and added terminal state transitions with zero reward to the training data.}

\begin{figure}[h]
    \centering
    \includegraphics[width=0.45\textwidth]{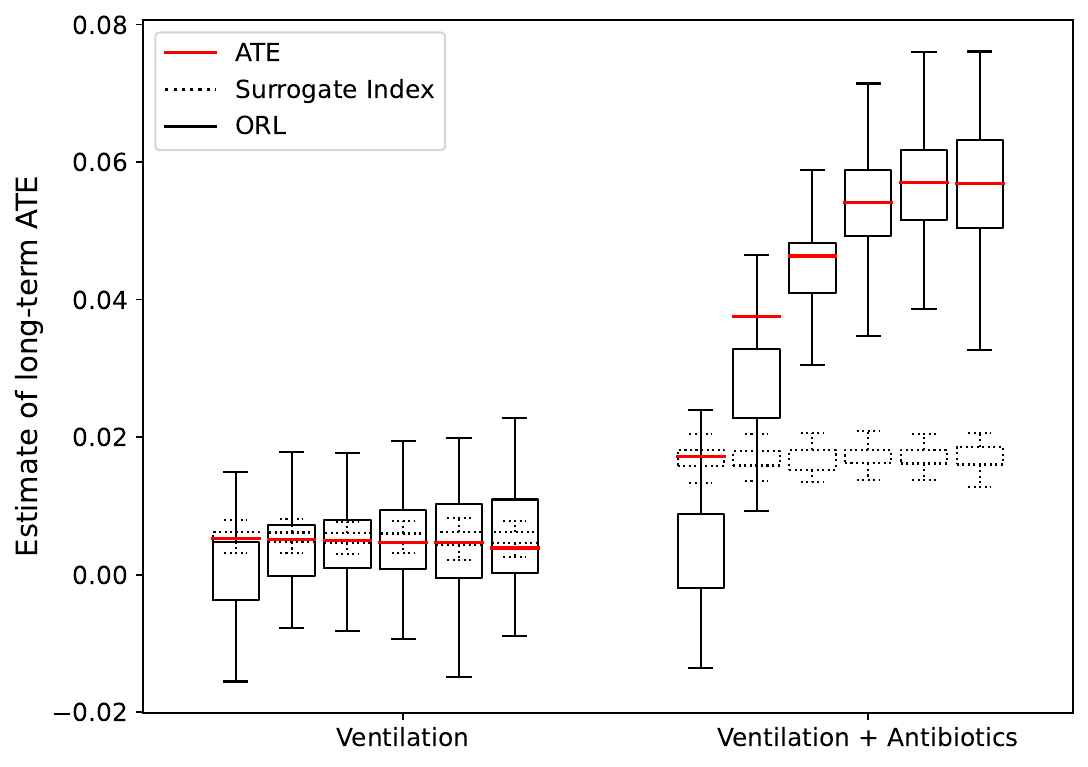} % Replace "filename" with your image file's name, without the extension
    \caption{Estimated ATEs of two treatments for Sepsis, for treatment durations of 1, 6, 12, 24, 48 and $\infty$}
    \label{fig:xp-sepsis}
\end{figure}

We mimic experiments by using each of the available medical treatments, the cartesian product of \{antibiotics, vassopressors, ventilation\}, as binary treatments to evaluate against the control of null treatment. As before, we simulate ground truth ATEs by applying the treatment for varying durations (1, 6, 12, 24, 48 and $\infty$) and then generate an experimental dataset of treatment and control. Results for two of the treatments, ventilation and ventilation in addition to antibiotics, spanning the spectrum of behavior of all 7 treatments, is shown in Figure \ref{fig:xp-sepsis}.

For both treatments, estimates for the ORL method largely center on the true ATE. Estimates of the ATE from ventilation are close to zero and correspondingly confidence intervals for all treatment durations span zero. On the other hand, treatment effects from ventilation and antibiotics together are estimated to be large and increasing with the duration of treatment, matching the increase in the true ATE with treatment duration. Interestingly, estimates for both treatments appear to degrade as the treatment duration shortens. This is due to the fact that the stationary policy used to mimic a short term experiment is an approximation that gets worse the shorter the target duration of treatment.

In addition, Figure \ref{fig:xp-sepsis} shows estimates for the surrogate index method where the surrogate variable is the state at $t=1$. As in the case for the simple MDP, the surrogate index estimate is only accurate for the treatment duration of 1 period. In the case of the ventilation treatment, the surrogate index estimate is correct but purely due to the fact that treatment duration has no effect on the long-term ATE. On the other hand, the ventilation and antibiotics estimates make it clear that the surrogate index method is biased when treatment duration has a sizeable effect on the long-term ATE.

\section{Conclusion}\label{sec-conclusion}
We develop a method for inferring the ATE of continual exposure to a long-term treatment, when only data from a short-term experiment is available. The key difficulty is that the treatment we consider is both novel and long-term. Together, this means that surrogate methods are unsuitable since surrogacy assumptions do not hold. Instead, we proceed by making a connection to offline reinforcement learning and embed our problem within a Markov decision process.

We reframe the problem of estimating the long-term ATE as evaluating the difference in the long-term outcomes of two different policies: a treatment and control policy. By constructing stationary policies equivalent to arbitrary-duration treatment regimes, we are able to make use of tools from off-policy reinforcement learning. In particular, we use an estimator which depends on the $Q$ and density ratio functions. Importantly, the estimator is doubly-robust with respect to these nuisance functions and asymptotically efficient.

Experiments on simulated data demonstrate the effectiveness of the ORL method over alternative methods. Estimates from the ORL method generally match the true ATE for the full spectrum of treatment durations, from single-period to permanent, using only two periods of data from the an experiment. 

%In contrast, the surrogate index method matches the true ATE only when surrogates are observed up to or beyond the point where the treatment is last active. Hence for long-term treatments, the measurement requirements of the surrogate index method are severe. 

In conclusion, our proposed method provides a robust and efficient solution for estimating the long-term ATE of continual exposure to a novel long-term treatment, when only short-term experimental data are available. %The approach can be seen as a complement to surrogate methods, which are well suited to short-term treatments. 
We do so without the need for long-term data, thereby bridging a significant gap in the study of long-term treatments. This opens up new possibilities for research and interventions in various fields where understanding the long-term effects of treatments is crucial.

\section*{Impact statement}
This paper presents work whose goal is to advance the field of Machine Learning. There are many potential societal consequences of our work, none which we feel must be specifically highlighted here.

\bibliography{Refs}

\begin{thebibliography}{23}
\providecommand{\natexlab}[1]{#1}
\providecommand{\url}[1]{\texttt{#1}}
\expandafter\ifx\csname urlstyle\endcsname\relax
  \providecommand{\doi}[1]{doi: #1}\else
  \providecommand{\doi}{doi: \begingroup \urlstyle{rm}\Url}\fi

\bibitem[Athey et~al.(2019)Athey, Chetty, Imbens, and Kang]{athey_index}
Athey, S., Chetty, R., Imbens, G.~W., and Kang, H.
\newblock {The Surrogate Index: Combining Short-Term Proxies to Estimate
  Long-Term Treatment Effects More Rapidly and Precisely}.
\newblock NBER Working Papers 26463, National Bureau of Economic Research, Inc,
  November 2019.
\newblock URL \url{https://ideas.repec.org/p/nbr/nberwo/26463.html}.

\bibitem[Athey et~al.(2020)Athey, Chetty, and Imbens]{athey2020combining}
Athey, S., Chetty, R., and Imbens, G.
\newblock Combining experimental and observational data to estimate treatment
  effects on long term outcomes.
\newblock \emph{arXiv preprint arXiv:2006.09676}, 2020.

\bibitem[Battocchi et~al.(2021)Battocchi, Dillon, Hei, Lewis, Oprescu, and
  Syrgkanis]{battocchi}
Battocchi, K., Dillon, E., Hei, M., Lewis, G., Oprescu, M., and Syrgkanis, V.
\newblock Estimating the long-term effects of novel treatments.
\newblock In Ranzato, M., Beygelzimer, A., Dauphin, Y., Liang, P., and Vaughan,
  J.~W. (eds.), \emph{Advances in Neural Information Processing Systems},
  volume~34, pp.\  2925--2935. Curran Associates, Inc., 2021.

\bibitem[Bertsekas(2001)]{bertsekas-dynamicProgramming}
Bertsekas, D.
\newblock \emph{Dynamic Programming and Optimal Control}, volume 1 and 2.
\newblock Athena Scientific, 2 edition, 2001.

\bibitem[Bica et~al.(2020)Bica, Alaa, and Van Der~Schaar]{bica2020}
Bica, I., Alaa, A.~M., and Van Der~Schaar, M.
\newblock Time series deconfounder: estimating treatment effects over time in
  the presence of hidden confounders.
\newblock In \emph{Proceedings of the 37th International Conference on Machine
  Learning}, ICML'20. JMLR.org, 2020.

\bibitem[Chen et~al.(2007)Chen, Geng, and Jia]{chen2007}
Chen, H., Geng, Z., and Jia, J.
\newblock Criteria for surrogate end points.
\newblock \emph{Journal of the Royal Statistical Society. Series B (Statistical
  Methodology)}, 69\penalty0 (5):\penalty0 919--932, 2007.
\newblock ISSN 13697412, 14679868.
\newblock URL \url{http://www.jstor.org/stable/4623303}.

\bibitem[Chernozhukov et~al.(2018)Chernozhukov, Chetverikov, Demirer, Duflo,
  Hansen, Newey, and Robins]{cherno_dml}
Chernozhukov, V., Chetverikov, D., Demirer, M., Duflo, E., Hansen, C., Newey,
  W., and Robins, J.
\newblock {Double/debiased machine learning for treatment and structural
  parameters}.
\newblock \emph{The Econometrics Journal}, 21\penalty0 (1):\penalty0 C1--C68,
  01 2018.
\newblock ISSN 1368-4221.
\newblock \doi{10.1111/ectj.12097}.
\newblock URL \url{https://doi.org/10.1111/ectj.12097}.

\bibitem[Chernozhukov et~al.(2023)Chernozhukov, Newey, Singh, and
  Syrgkanis]{chernozhukov2023automatic}
Chernozhukov, V., Newey, W., Singh, R., and Syrgkanis, V.
\newblock Automatic debiased machine learning for dynamic treatment effects and
  general nested functionals, 2023.

\bibitem[Huang et~al.(2023)Huang, Wang, Yuan, Zhao, and Zhang]{huang-ltt2023}
Huang, S., Wang, C., Yuan, Y., Zhao, J., and Zhang, J.
\newblock Estimating effects of long-term treatments.
\newblock In \emph{Proceedings of the 24th ACM Conference on Economics and
  Computation}, EC '23, pp.\  907, New York, NY, USA, 2023. Association for
  Computing Machinery.
\newblock ISBN 9798400701047.
\newblock \doi{10.1145/3580507.3597701}.
\newblock URL \url{https://doi.org/10.1145/3580507.3597701}.

\bibitem[Imbens et~al.(2023)Imbens, Kallus, Mao, and Wang]{imbens2023longterm}
Imbens, G., Kallus, N., Mao, X., and Wang, Y.
\newblock Long-term causal inference under persistent confounding via data
  combination, 2023.

\bibitem[Kallus \& Mao(2020)Kallus and Mao]{kallus2020role}
Kallus, N. and Mao, X.
\newblock On the role of surrogates in the efficient estimation of treatment
  effects with limited outcome data.
\newblock \emph{arXiv preprint arXiv:2003.12408}, 2020.

\bibitem[Kallus \& Uehara(2022)Kallus and Uehara]{kallus_uehara}
Kallus, N. and Uehara, M.
\newblock Efficiently breaking the curse of horizon in off-policy evaluation
  with double reinforcement learning.
\newblock \emph{Operations Research}, 70\penalty0 (6):\penalty0 3282--3302,
  2022.
\newblock \doi{10.1287/opre.2021.2249}.
\newblock URL \url{https://doi.org/10.1287/opre.2021.2249}.

\bibitem[Kennedy(2023)]{kennedy2023semiparametric}
Kennedy, E.~H.
\newblock Semiparametric doubly robust targeted double machine learning: a
  review, 2023.

\bibitem[Lewis \& Syrgkanis(2021)Lewis and Syrgkanis]{lewis2021}
Lewis, G. and Syrgkanis, V.
\newblock Double/debiased machine learning for dynamic treatment effects.
\newblock In Ranzato, M., Beygelzimer, A., Dauphin, Y., Liang, P., and Vaughan,
  J.~W. (eds.), \emph{Advances in Neural Information Processing Systems},
  volume~34, pp.\  22695--22707. Curran Associates, Inc., 2021.
\newblock URL
  \url{https://proceedings.neurips.cc/paper_files/paper/2021/file/bf65417dcecc7f2b0006e1f5793b7143-Paper.pdf}.

\bibitem[Liu et~al.(2018)Liu, Li, Tang, and Zhou]{liuNEURIPS2018}
Liu, Q., Li, L., Tang, Z., and Zhou, D.
\newblock Breaking the curse of horizon: Infinite-horizon off-policy
  estimation.
\newblock In Bengio, S., Wallach, H., Larochelle, H., Grauman, K.,
  Cesa-Bianchi, N., and Garnett, R. (eds.), \emph{Advances in Neural
  Information Processing Systems}, volume~31. Curran Associates, Inc., 2018.
\newblock URL
  \url{https://proceedings.neurips.cc/paper_files/paper/2018/file/dda04f9d634145a9c68d5dfe53b21272-Paper.pdf}.

\bibitem[Mnih et~al.(2015)Mnih, Kavukcuoglu, Silver, Rusu, Veness, Bellemare,
  Graves, Riedmiller, Fidjeland, Ostrovski, Petersen, Beattie, Sadik,
  Antonoglou, King, Kumaran, Wierstra, Legg, and Hassabis]{mnih2015humanlevel}
Mnih, V., Kavukcuoglu, K., Silver, D., Rusu, A.~A., Veness, J., Bellemare,
  M.~G., Graves, A., Riedmiller, M., Fidjeland, A.~K., Ostrovski, G., Petersen,
  S., Beattie, C., Sadik, A., Antonoglou, I., King, H., Kumaran, D., Wierstra,
  D., Legg, S., and Hassabis, D.
\newblock Human-level control through deep reinforcement learning.
\newblock \emph{Nature}, 518\penalty0 (7540):\penalty0 529--533, February 2015.
\newblock ISSN 00280836.
\newblock URL \url{http://dx.doi.org/10.1038/nature14236}.

\bibitem[Murphy(2003)]{murphy2003}
Murphy, S.~A.
\newblock Optimal dynamic treatment regimes.
\newblock \emph{Journal of the Royal Statistical Society: Series B (Statistical
  Methodology)}, 65\penalty0 (2):\penalty0 331--355, 2003.
\newblock \doi{https://doi.org/10.1111/1467-9868.00389}.
\newblock URL
  \url{https://rss.onlinelibrary.wiley.com/doi/abs/10.1111/1467-9868.00389}.

\bibitem[Oberst \& Sontag(2019)Oberst and Sontag]{pmlr-v97-oberst19a}
Oberst, M. and Sontag, D.
\newblock Counterfactual off-policy evaluation with {G}umbel-max structural
  causal models.
\newblock In Chaudhuri, K. and Salakhutdinov, R. (eds.), \emph{Proceedings of
  the 36th International Conference on Machine Learning}, volume~97 of
  \emph{Proceedings of Machine Learning Research}, pp.\  4881--4890. PMLR,
  09--15 Jun 2019.
\newblock URL \url{https://proceedings.mlr.press/v97/oberst19a.html}.

\bibitem[Prentice(1989)]{prentice1989}
Prentice, R.~L.
\newblock Surrogate endpoints in clinical trials: Definition and operational
  criteria.
\newblock \emph{Statistics in Medicine}, 8\penalty0 (4):\penalty0 431--440,
  1989.
\newblock \doi{https://doi.org/10.1002/sim.4780080407}.
\newblock URL
  \url{https://onlinelibrary.wiley.com/doi/abs/10.1002/sim.4780080407}.

\bibitem[Sutton \& Barto(1998)Sutton and Barto]{sutton1998reinforcement}
Sutton, R.~S. and Barto, A.~G.
\newblock \emph{Reinforcement Learning: An Introduction}.
\newblock MIT Press, Cambridge, MA, 1998.

\bibitem[Uehara et~al.(2020)Uehara, Huang, and Jiang]{ueharaminimax}
Uehara, M., Huang, J., and Jiang, N.
\newblock Minimax weight and q-function learning for off-policy evaluation.
\newblock In \emph{Proceedings of the 37th International Conference on Machine
  Learning}, ICML'20. JMLR.org, 2020.

\bibitem[VanderWeele(2013)]{vanderweele}
VanderWeele, T.~J.
\newblock Surrogate measures and consistent surrogates.
\newblock \emph{Biometrics}, 69\penalty0 (3):\penalty0 561--569, 2013.
\newblock ISSN 0006341X, 15410420.
\newblock URL \url{http://www.jstor.org/stable/24538119}.

\bibitem[Yang et~al.(2023)Yang, Eckles, Dhillon, and Aral]{yang2023targeting}
Yang, J., Eckles, D., Dhillon, P., and Aral, S.
\newblock Targeting for long-term outcomes.
\newblock \emph{Management Science}, 0\penalty0 (0), 2023.

\end{thebibliography}
\bibliographystyle{icml2024}
\newpage

\appendix
\onecolumn
\section{Proofs}
\subsection{Proof of Theorem 1}
\label{sec:proof_ident}
Since expectations are linear, it suffices to show that each per period outcome of the long-term outcome (each term in Equation \eqref{eq:summable_ate}) can be expressed as a function of observable data. For periods beyond the first:
\begin{align}
    \mathbb{E} \left[Y_t(\pi^T) |S_0, A_0 \right] &= \mathbb{E}_Y\left[Y_t(\pi^T) \mid S_0, A_0 \right] \nonumber \\
    &= \mathbb{E}_{Y}\left[Y_t(\pi^T) \mid \pi^T, S_0, A_0 \right] \nonumber \\
    &= \mathbb{E}_{Y}\left[Y_t \mid A_t=1_{t<T}, \pi^T, S_0, A_0 \right] \nonumber \\
    &= \mathbb{E}_{S_t}\left[\mathbb{E}_{Y}\left[Y_t \mid A_t=1_{t<T}, S_t \right] | \pi^T, S_0, A_0 \right]\nonumber \\
    &=  \mathbb{E}_{S}\left[\mathbb{E}_{Y}\left[Y \mid A=1_{t<T}, S  \right] \mid \pi^T, S_0, A_0;t \right] 
    \label{eq:identification}
\end{align}
The first and fourth equalities rely on the law of iterated expectations, the second is justified via unconfoundedness and the third uses the definition of a potential outcome. The final equality relies on stationarity where the notation $\mathbb{E}\left[\cdot ; t \right]$ denotes the expectation induced by projecting $t$ periods ahead under the Markov model. The same derivation can be done for the first period where the action is $A_0$.

Applying this for all periods
\begin{equation}
     \mathbb{E}\left[ \sum_t \gamma^t Y_t(\pi^T)  |S_0, A_0\right] = 
     \mathbb{E}_{Y}\left[Y \mid A_0, S_0  \right]
     \sum_{t=1} \gamma^t \sum_{S}\mathbb{E}_{Y}\left[Y \mid A=1_{t<T}, S  \right] p(S|\pi^T, S_0, A_0; t).
\end{equation}
From here, one can use the definition of the $Q$ function and use the standard proof to show the equivalence of expected discounted rewards from an initial state-action to the $Q$ function \cite{sutton1998reinforcement}.

\section{Details of Experiments}
\subsection{Q function and density ration model implementation details}

The ORL method requires the estimation of two nuisance functions, the $Q$ function and the density ratio functions. For the $Q$ function, we use a feed-forward neural network parameterized separately for each of treatment and control. Each network consists of two hidden layers with 128 and 64 features respectively with sigmoid activation functions and a linear final layer with no activation function. Additionally, we maintain separate ``target'' networks by freezing the parameters of each network for 64 epochs, which proved invaluable in stabilizing training \cite{mnih2015humanlevel}.

For the density ratio functions, we use the Minimax weight estimator from \cite{ueharaminimax} where we restrict both the discriminator and density ratio function classes to be linear with the feature maps $\phi(s, a) = \begin{bmatrix} s & s^2 & s\*a& s^2 &s^2 a & a & 1 \end{bmatrix}$.

\subsection{Sepsis simulator model details}

As the sepsis simulator relies on discrete states, with enough data, a one-hot encoded representation of the states identifying each discrete state would recover the ATE exactly. To ensure a realistic setting where the true states are unknown, we keep states in the original human-interpretable format, with each vital sign and its measurement as a dimension of the state-space. In addition, the simulator generates trajectories that are episodic since death and discharge are terminal events whereas our method applies to tasks that are continuing. To handle this, we force outputs from the learned $Q$ function to be zero for terminal states. Alternatively, we could have introduced another dimension of the state-space to be a boolean terminal state indicator and added terminal state transitions with zero reward to the training data.

\begin{table}
\centering
  \caption{Simulation Parameter Values}
  \label{tab:freq}
  \begin{tabular}{ccl}
    \toprule
    Notation&Description&Value\\
    \midrule
    $\gamma$ & Discount rate & 0.9 \\
    $\mu$ & Drift in state transition& 0.05\\
    $\sigma_s$ & Std. dev. in state transition in 5& 0.1\\
    $\sigma_r$ & Std. dev. in state-outcome mapping & 0.1\\
    $\theta $ & Curvature in state-outcome mapping & 0.8\\
    $\tau_s$ & Treatment effect on state transition& 0.05\\
    $\tau_y$ & Treatment effect on per-period outcome& 0.01\\
  \bottomrule
\end{tabular}
\end{table}

\end{document}